\documentclass[10pt,final,journal,a4paper,twoside,twocolumn,romanappendices]{IEEEtran}
\usepackage[tbtags]{amsmath}
\usepackage{amsmath}

\usepackage{algorithm}
\usepackage{algpseudocode}

\usepackage{amssymb}
\usepackage{cite}
\usepackage{stfloats}

\usepackage{amsthm}
\usepackage{float}
\usepackage{amsthm}
\usepackage{amssymb}

\usepackage{graphicx}%
\usepackage{acronym}
\usepackage[keeplastbox]{flushend}
\usepackage{stfloats}
\usepackage{color}
\usepackage{subcaption}

\newtheoremstyle{myremark}%
  {}
  {}
  {}
  {\parindent}
  {\itshape}
  {:}
  {5pt plus 1pt minus 1pt}
  {\thmname{#1}\thmnumber{~#2}\thmnote{~(#3)}}

\theoremstyle{myremark}

\theoremstyle{myremark}

\acrodef{CRN}{cognitive radio network}
\acrodef{SU}{secondary user}
\acrodef{PU}{primary user}
\acrodef{ZF}{zero forcing}
\acrodef{FD}{full-duplex}
\acrodef{BS}{base station}
\acrodef{i.i.d.}{independent and identically distributed}
\acrodef{DL}{downlink}
\acrodef{UL}{uplink}
\acrodef{SINR}{signal-to-interference-plus-noise ratio}
\acrodef{SNR}{signal noise ratio}
\acrodef{AWGN}{additive white Gaussian noise}
\acrodef{MMSE}{minimum mean square error}
\acrodef{SIC}{successive interference cancellation}
\acrodef{SI}{self-interference}
\acrodef{CCI}{co-channel interference}
\acrodef{MUI}{multiuser interference}
\acrodef{NOMA}{non-orthogonal multiple access}
\acrodef{OMA}{orthogonal multiple access}
\acrodef{QoS}{quality of service}
\acrodef{SIC}{successive interference cancellation}
\acrodef{SVD}{singular value decomposition}
\acrodef{MIMO}{multiple-input multiple-output}
\acrodef{SISO}{single-input single-output}
\acrodef{MIMO-NOMA}{multiple-input multiple-output non-orthogonal multiple access}
\acrodef{MIMO-OMA}{multiple-input multiple-output orthogonal multiple access}

\begin{document}
%
\title{Resource Allocation for Downlink NOMA Systems: Key Techniques and Open Issues}

\author{\IEEEauthorblockN{S. M. Riazul Islam, Ming Zeng, Octavia A. Dobre and Kyung-Sup Kwak
}

}

\maketitle


\begin{abstract}
This article presents advances in resource allocation (RA) for downlink non-orthogonal multiple access (NOMA) systems, focusing on user pairing (UP) and power allocation (PA) algorithms. The former pairs the users to obtain the high capacity gain by exploiting the channel gain difference between the users, while the later allocates power to users in each cluster to balance system throughput and user fairness. Additionally, the article introduces the concept of cluster fairness and proposes the divide-and-next largest difference-based UP algorithm to distribute the capacity gain among the NOMA clusters in a controlled manner. Furthermore, {\color{black}performance comparison between multiple-input multiple-output NOMA (MIMO-NOMA) and MIMO-OMA is conducted when users have pre-defined quality of service. Simulation results are presented, which validate the advantages of NOMA over OMA.} Finally, the article provides avenues for further research on RA for downlink NOMA.
\end{abstract}

\begin{IEEEkeywords}
5G, NOMA, Resource allocation, User pairing, Power allocation.
{\let\thefootnote\relax\footnote{This research was supported by the Ministry of Science, ICT and Future Planning (MSIP), Korea, as well as the Natural Sciences and Engineering Research Council of Canada (NSERC). 

S. M. R. Islam (e-mail: riaz@sejong.ac.kr) is with the Department of Computer Science and Engineering, Sejong University, South Korea.

M. Zeng (e-mail: mzeng@mun.ca) and O. A. Dobre (e-mail: odobre@mun.ca) are with the Faculty of Engineering and Applied Science, Memorial University, Canada.

K. S. Kwak (e-mail: kskwak@inha.ac.kr) is with the UWB Wireless Communications Research Center, Inha University, South Korea.
}  }
\end{IEEEkeywords}

%
\IEEEpeerreviewmaketitle

\section{Introduction}
Non-orthogonal multiple access (NOMA) enables a balanced tradeoff between spectral efficiency and user fairness, being recognized as a promising multiple access technique for the fifth generation (5G) networks \cite{3, 34, focus}. In contrast to orthogonal multiple access (OMA), NOMA exploits power domain to simultaneously serve multiple users at different power levels, where the power allocation (PA) for each user plays a key role in determining the overall performance of the system. Downlink NOMA combines superposition coding at the base station (BS) and successive interference cancellation (SIC) decoding at the user. To maintain user fairness, NOMA allocates more power to the users with weaker channel gains. Because of additional system overhead for channel feedback coordination and error propagation, it is not feasible to apply NOMA on all users jointly. Therefore, the idea of user pairing (UP) has emerged \cite{19}, with users in the cell divided into multiple clusters and NOMA is employed within each cluster (see Fig. 1). The performance of a NOMA system is highly dependent on both UP and PA. These are usually referred to as resource allocation (RA), which represents the central theme of this article. The RA in NOMA aims to determine the users to be paired and power to be allocated to each user within each cluster. The optimal performance of NOMA RA can be attained by an exhaustive search of all possible user pairs and transmit power allocations, which is, however, computationally complex. Moreover, if dynamic UP and PA are adopted, the decoding order in SIC and PA ratios introduce additional signaling overheads.

To date, extensive research has been performed on NOMA RA due to the pivotal role of the RA algorithms in achieving the benefits offered by NOMA. To this end, this article appraises the state-of-the-art of RA algorithms for downlink NOMA research and uncovers various issues to be addressed. More specifically, it

\begin{itemize}
\item provides a categorized survey of the UP {\color{black}and PA} algorithms;

\item proposes an UP algorithm which ensures the cluster fairness in terms of sum rate gain of desired degree;

\item {\color{black}conducts performance comparison between multiple-input multiple-output NOMA (MIMO-NOMA) and MIMO-OMA when users have pre-defined quality of service (QoS) requirements;}

\item highlights challenges and open issues to be addressed in downlink NOMA RA.
\end{itemize}

\begin{figure}
\centering
\includegraphics[width=0.5\textwidth]{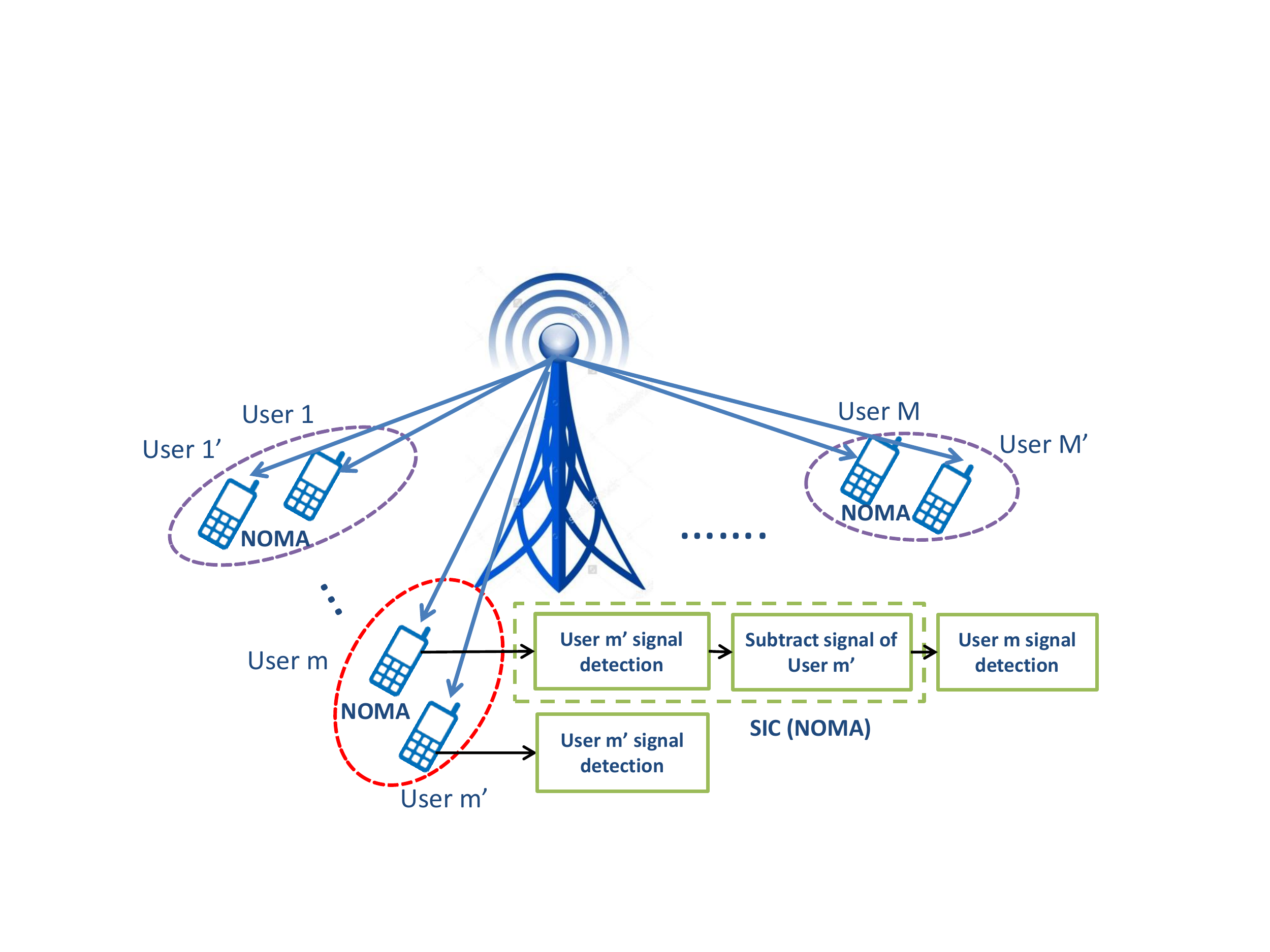}
\caption{A downlink NOMA system with multiple clusters.}
\end{figure} 

\section{User Paring in NOMA}
Based on the desired performance (e.g., sum rate gain), deployment environment, and {\color{black}implementation complexity}, there exist a number of UP algorithms. UP ideally should be compatible with the   PA strategy to provide high throughput with minimum computational complexity, while maintaining user fairness. UP algorithms for both single-input single-output (SISO) and MIMO-NOMA are introduced as follows. 

\subsection{UP in SISO-NOMA}
Random pairing is the easiest UP algorithm, in which the BS selects the users randomly from the set of candidates to form the clusters. Although it comes with the lowest complexity, it exhibits suboptimal sum rate performance because of not taking the users' channel gains into account. The mathematical investigation on UP shows that the performance gain of NOMA with fixed PA (F-PA) over OMA grows with increasing the difference between the channel gains of the users of interest \cite{19}. As such, pairing the user of highest channel gain with the user of lowest channel gain provides the best performance gain, whereas the user of second highest channel gain should be paired with the user of second lowest channel gain to obtain the second best performance gain, and so on. This algorithm is referred to as the next largest difference-based UP algorithm (NLUPA) and is one of the most common techniques \cite{19}. In contrast to the above approach, the cognitive radio (CR)-inspired NOMA pairs the user of highest channel gain with the user of second highest channel gain, the user of third highest channel gain with the user of fourth highest channel gain, and so on, since the $nth$ user is opportunistically served on the condition that the QoS of the $m$th user $(n>m)$ is guaranteed \cite{19}. The idea of such an UP algorithm can be referred to as next best diversity pairing.

UP can also play a role in canceling adjacent channel interferences by adopting the vertical UP concept when adjacent sub-channels are sequentially assigned to successive user pairs \cite{21}. In this scheme, the users in each cluster apply additional SIC to cancel the interferences from the previous clusters. However, this algorithm comes with further computational complexity due to additional SIC operations. {\color{black}The previous UP algorithms do not consider the realistic fact that there might not exist enough strong users. In such a situation, there are leftover weak users after each strong user is paired with its partner. A possible solution for such a problem is the adoption of a hybrid approach, where some users are left without pairing and are accessed via OMA. However, this deprives such users of the advantages offered by NOMA. Alternatively, NOMA can be implemented using the concept of virtual UP \cite{22}, where a frequency band can be shared by two weak users of similar channel gains and a strong user: half of the bandwidth can be shared by the strong user and a weak user, and half is used by the strong user with the other weak user}.

An UP technique should be of low computational complexity. A good practice is to formulate the pre-requisites for UP such that some user groups which are not appropriate for NOMA multiplexing can be excluded from unnecessary comparison of candidate user pairs \cite{23}. The complexity and signaling overhead can also be reduced by using a pre-defined user grouping, where users are divided into different groups based on their channel conditions. Then, a pair can be formed only if the users come from different groups. Eventually, the BS does not need to convey the SIC order information to users in every sub-frame. The scheduled times, along with channel conditions, can also be determining factors for excluding users from the set of candidates. In such a case, the users in the cell are divided into two groups, with users in the first group having higher channel gains than users in the other group. Then, based on the scheduled times and signal-to-interference-plus-noise ratio (SINR), users with low scheduling priority can be excluded from each group by setting an SINR threshold. Finally, the BS finds the user with an optimum fairness metric in each group to form a pair.

{\color{black}
UP algorithms discussed above are generally applicable to NOMA systems where user grouping is likely to be in a partition form. However, there might arise some cases where merge-and-split-like algorithms cannot be applied \cite{34}. In such cases, a game theory framework could be useful. \cite{30} considers users and subchannels as two sets of players to be matched with each other and applies matching games to achieve the maximum weighted sum-rate. Note that [7] proposes many-to-many matching algorithms to perform UP for a more general NOMA system, where multiple users share each sub-channel and multiple sub-channels can be accessed by each user. Also, [10] investigates UP for a varying number of users to be multiplexed on a subcarrier.
}

\begin{figure}
\centering
\includegraphics[width=0.5\textwidth]{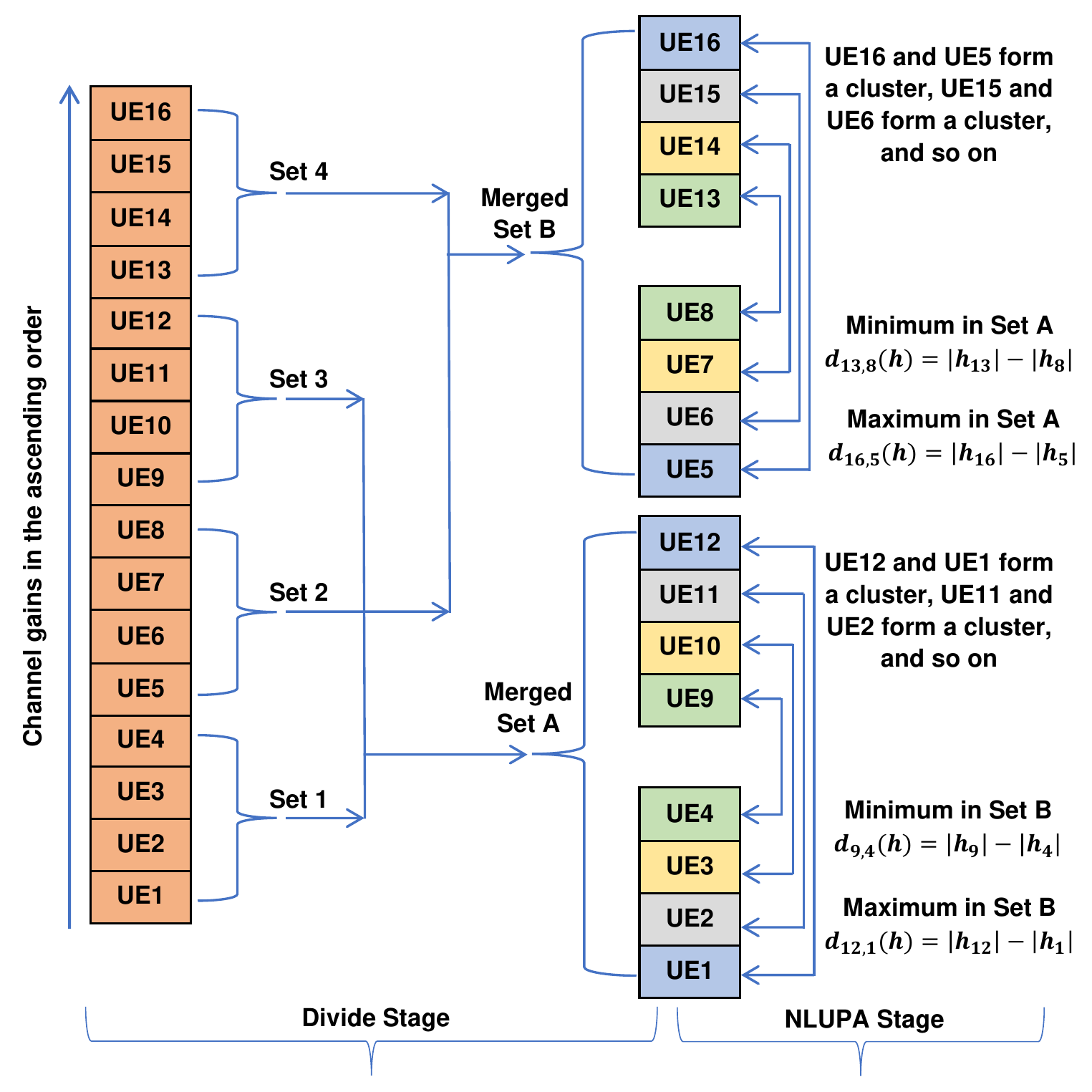}
\caption{Illustration of the D-NLUPA for a two-user NOMA {\color{black}$(N=16)$}. UE stands for user. }\vspace{-0.5cm}
\end{figure}

\vspace{-0.2cm}
\subsection{UP in MIMO-NOMA}
The UP in MIMO-NOMA also has impact on the sum rate gain. Interestingly, in a two-user cluster with zero-forcing precoding, the user with weak channel conditions has no influence on the strong user's rate. Therefore, the strong user could be first selected to form a cluster. Then, the weak user should be chosen to optimize the performance metric. The said optimization can usually be achieved with the minimization of the intra- and inter-cluster interferences. These interferences can be reduced if a cluster is formed by two users whose channel gains are sufficiently distinct but there exists a high correlation between their channels. The large-gain differences make certain the effectiveness of NOMA, whereas the high correlation helps eliminate the inter-cluster interference. Apart from zero-forcing precoding, the quasi-degrading\footnote[1]{For a particular decoding and encoding order $(n,m)$ of NOMA and dirty paper coding (DPC), respectively, two channel coefficients $h_m$ and $h_n$ are quasi-degraded with respect to the targeted SNR levels of the corresponding users if and only if the minimum transmission powers of NOMA and DPC are comparable \cite{24}. The idea of the said comparison with the DPC comes from the fact that the use of DPC can achieve the capacity region of the downlink broadcast channel with perfect channel state information (CSI) information available at the BS.} nature of NOMA channels is exploited to formulate a new form of precoding which is referred as the quasi-degraded channel-driven (QDC) precoding. With QDC precoding, a low complexity sequential UP algorithm is formulated to reduce the transmit power \cite{24}. However, this is not efficient when the number of transmit antennas at the BS is greater than the number of downlink users. To solve this problem, there exist a couple of variations, namely projection-based UP algorithm and inversion-based UP algorithm. Both zero-forcing and QDC precoding-based MIMO-NOMA deal with sum rate maximization problems. The minimum Euclidean distance precoding-based MIMO-NOMA, on the other hand, focuses on the symbol error rate (SER) reduction. Based on this precoding, a pair of UP algorithms are proposed to further reduce the SER, namely, the condition number-based and orthogonality defect-based UP algorithms, in which the basic ideas are that two users with smaller condition number and smaller orthogonality defect should be paired, respectively \cite{28}.

\vspace{-0.2cm}
\subsection{The Concept of Divide-and-NLUPA}
Here, we propose a modified NLUPA scheme, referred to as divide-and-NLUPA (D-NLUPA), to guarantee a minimum sum rate gain for each cluster, {\color{black}and thereby, introduce the concept of cluster fairness}. NLUPA, as described in Section III. A, pairs the user of best channel gain with the user of worst channel gain, the user of second best channel gain with the user of second lowest channel gain, and so on. As the sum rate gain achieved by NOMA when compared with OMA is logarithmically proportional to the ratio of the channel gains of the strong and weak users in a cluster, the difference between the channel gains  affects the performance gain. {\color{black}In this paper, the difference between the orders of the paired strong and weak users is called the range, i.e., $|n-m|$, where $n$ and $m$ denote the orders of the strong and weak users, respectively. The corresponding channel gain difference is referred to as the distance, i.e., $d_{n,m} (h)=|h_n |-|h_m |$, where $h_n$ and $h_m$ denote the channel gains of the strong and weak users, respectively. As the range increases, the distance grows as well, yielding a higher performance gain}. Assume the total number of users is $N$.\footnote[2]{Without loss of generality, we assume that $N$ is an even number. If $N$ is odd, different UP strategies can be applied, as presented in Section III. A.} Thus, the first cluster enjoys the maximum gain, while the gain for the $\frac{N}{2}$th cluster may not be significant. {\color{black}Therefore, it is possible to obtain some clusters which actually do not enjoy any sum-rate gain (gain is very closed to zero)}. 

{\color{black}To guarantee a minimum performance gain, we introduce the "divide" step in D-NLUPA, by setting a minimum range, and thus, increasing the corresponding minimum value of the distance}. Because of this "divide" step, the scenario of near-zero gain clusters can be avoided and each cluster enjoys a minimum gain as designed. The value of this {\color{black}minimum gain} depends on how we shuffle the users in the "divide" stage. To illustrate the idea, an example is shown in Fig. 2, with $N=16$. {\color{black}The ordered users are first divided into $\frac{N}{z}=4$ sets, where $z=4$ is the number of users in each set. Then, sets 1 and 3 are merged into set A, sets 2 and 4 are merged into set B}, and finally NLUPA is respectively applied to sets A and B to form the clusters; this yields {\color{black}the minimum value of the range} {\color{black}and the corresponding minimum value of the distance in each set. Note that although the minimum values of the range in sets A and B are the same, the corresponding minimum values of the distance are not necessarily the same due to different channel gains of the users in two sets. The minimum distance, $d_{n,m}(h), n,m \in \{1,\cdots,16\}$ is considered over both sets. }


To compare the sum rate gain of NLUPA and D-NLUPA, Fig. 3(a) presents results from simulation. It can be noticed that this gain has been controlled by {\color{black}changing} the minimum distance in D-NLUPA, while there is no such control in NLUPA. {\color{black}The "divide" stage in D-NLUPA arranges the user pairing in such a way that there occurs a certain minimum distance (here, around 15 dB) between the two users forming a cluster. Because of this minimum distance, D-NLUPA guarantees a minimum sum rate gain for each cluster.} Let us now merge and sort the sum rate gains of the clusters formed from both A and B sets. Then, the sum rate gain vs. cluster index performance is presented in Fig. 3(b). Results show that about $50\%$ D-NLUPA-based clusters exhibit higher gains compared with NLUPA, while the remaining DNLUPA-based clusters obtain lower gains. The advantage of the cluster fairness is thus achieved by redistributing the gains to the clusters in a controlled manner, while the aggregated throughput remains the same for both NLUPA and D-NLUPA. {\color{black}Also, it is noticeable that random pairing obtains the lowest sum rate gain.}

\begin{figure*}
\centering
\begin{subfigure}{0.5\textwidth}
  \centering
  \includegraphics[width=1\linewidth]{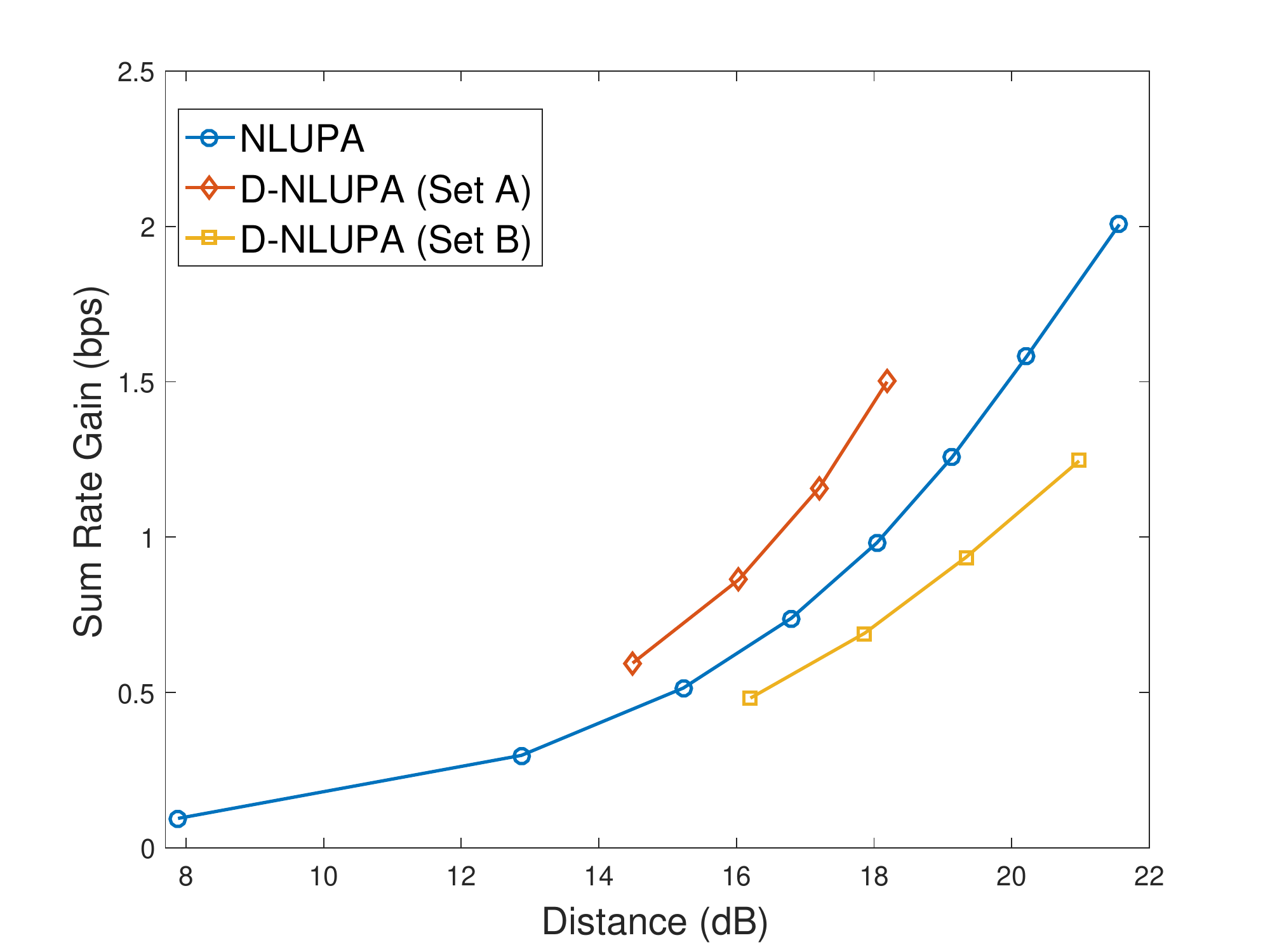}
  \caption{}
  \label{fig:sub1}
\end{subfigure}%
\begin{subfigure}{0.5\textwidth}
  \centering
  \includegraphics[width=1\linewidth]{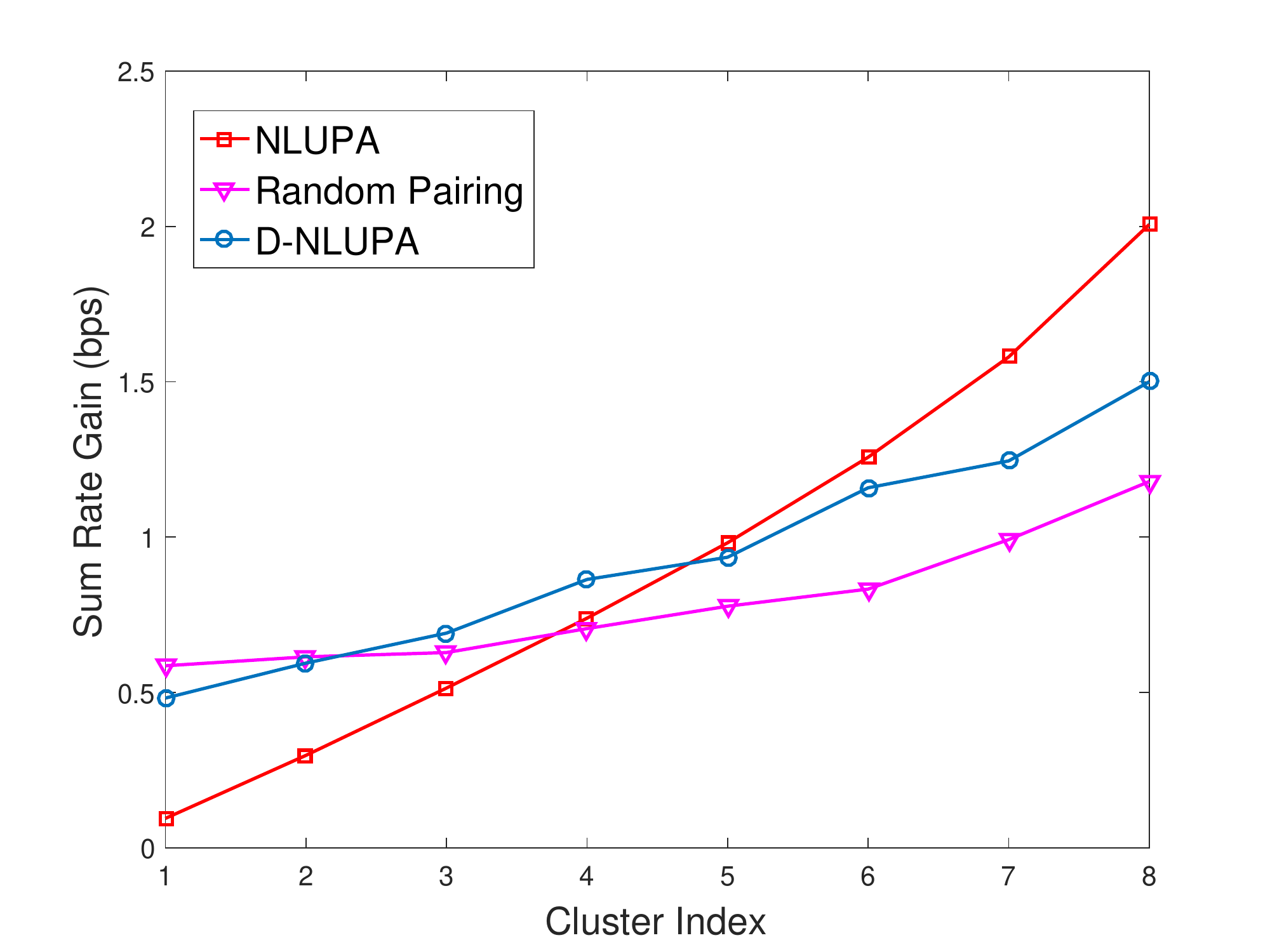}
  \caption{}
  \label{fig:sub2}
\end{subfigure}
\caption{{\color{black}NLUPA and D-NLUPA with F-PA: Sum rate gain versus a) distance and b) cluster index. Transmit power is 1 W; PA ratio is 0.6:0.4; cell radius is 100 m; $N=16$; Rayleigh fading channel with log-normal shadowing of path loss exponent of 2.5 and standard deviation of 3 is considered.}}
\label{fig:test}
\end{figure*}

\section{Power Allocation in NOMA}
Compared with OMA, the role of PA in NOMA is further enhanced,
since users are multiplexed in the power domain. Interference management, rate distribution,
and even user admission are directly impacted by PA. Generally, PA in NOMA is determined
by the users' channel conditions, availability of CSI, QoS requirements, total power constraint and system objective.
An inappropriate PA not only leads to an unfair rate distribution among users, but also causes
system outage as SIC may fail. There are different PA performance metrics, e.g., the number
of admitted users, sum rate, user fairness, outage probability and total power consumption. Thus, PA in NOMA
should aim at achieving either more admitted users and higher sum rate, or a balanced fairness
under minimum power consumption. A variety of PA strategies have been proposed in the literature, targeting
different aspects of PA in NOMA, and a classification is provided in Fig. 4. We introduce PA in the following two subsections: one focuses on single-carrier (SC) SISO  systems,\footnote[3]{In the following sections, we will simply use SISO to refer to SC SISO. Further, MC-NOMA and MIMO-NOMA refer to MC SISO-NOMA and SC MIMO-NOMA, respectively.} while the other deals with
multi-carrier (MC) and MIMO systems, respectively.

\begin{figure}
\centering
\includegraphics[width=0.5\textwidth]{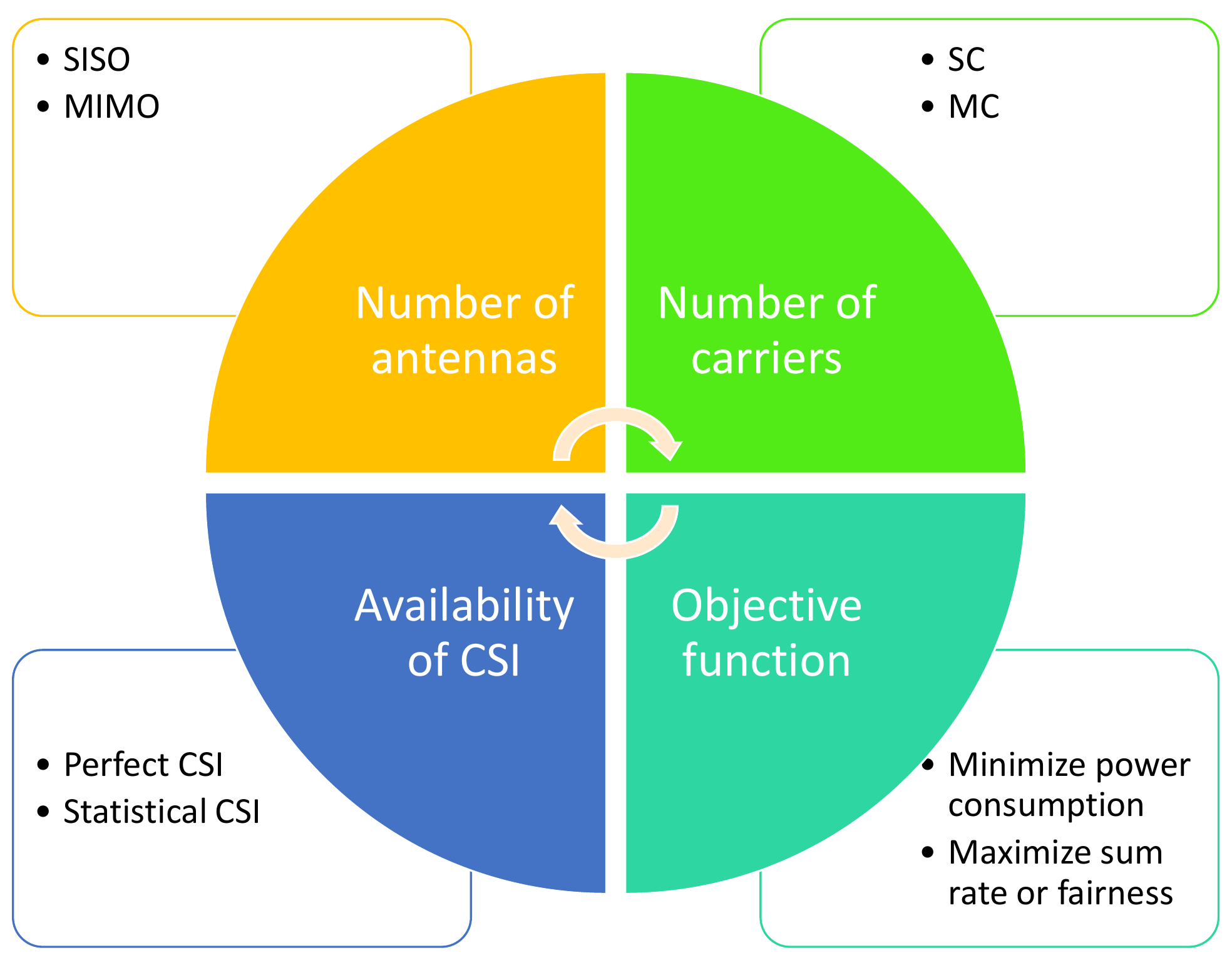}
\caption{Classification of PA strategies. }
\end{figure} 

\subsection{PA in SISO-NOMA}
Early works in PA mainly target SISO systems. Unlike OMA, whose optimal PA to maximize
the sum rate follows the water-filling technique, the corresponding PA for NOMA simply allocates
all power to the user with the best channel \cite{31, EE}. Obviously, this leads to extreme unfairness among
users and also diminishes the number of admitted users. To balance system throughput and user
fairness, NOMA allocates more power to the weak user. This way, the strong user handles the interference from the weak user using SIC, while the interference to its counterpart remains
comparatively small. {\color{black}The simplest PA algorithm is F-PA, which allocates power to each user using a fixed ratio based on its position in the channel ordering. Since the users' specific channel gains are not considered during PA, F-PA cannot satisfy users' various QoS requirements. To address this issue,} fractional transmit power control (FTPC) allocates power to each user inversely proportional to its channel gain powered with a decaying factor. {\color{black}Nevertheless,} assigning the same decaying factor to all users is still suboptimal, and selecting the appropriate decaying factor to balance system throughput and user fairness remains an open issue. 


{\color{black}PA is directly impacted by the availability of CSI. Under perfect CSI, the multi-user weighted sum rate maximization problem is proved to be convex, and thus optimal PA can be obtained using convex optimization \cite{31}. The max-min fairness problem is shown to be quasi-convex, and optimal PA can be obtained via the bisection method \cite{13}. The energy-efficient PA problem is formulated as a difference of two convex functions, and PA can be obtained by iteratively solving the convex sub-problems \cite{32}. Under statistical CSI, although the min-max outage probability under a given SIC order is shown to be non-convex, optimal PA is derived in \cite{13}; on this basis, \cite{29} further obtains the corresponding optimal SIC decoding order.
}

Note that the works above do not ensure a higher throughput of NOMA over OMA. To achieve this for the weak user, CR-inspired PA can be applied, where NOMA is considered as a special case of CR networks and the weak user is viewed as a primary user \cite{19}. However, this may sacrifice the performance of the strong user since it is served only after the weak user's QoS is met.
To overcome this issue, dynamic PA is proposed in \cite{17}, which allocates power to users such
that the individual user rate achieved by NOMA is strictly larger than that provided by OMA.


\vspace{-0.2cm}
\subsection{PA in MC-NOMA and MIMO-NOMA}
For multi-user systems, NOMA is usually integrated with MC (MC-NOMA) to reduce
complexity. In MC-NOMA, {\color{black}a user can occupy multiple sub-carriers, and vice versa.} MC-NOMA is quite suitable for 5G as it is difficult
to find continuous wide bandwidth in 5G. Compared with orthogonal frequency-division multiple access (OFDMA), MC-NOMA can further increase spectral efficiency and the number of {\color{black}simultaneously supported} users. Its performance depends on both PA and sub-carrier assignment (SA). Some works maximize the sum rate under total power constraint to increase spectral efficiency and throughput, whereas others minimize the total power consumption under QoS requirements to improve energy efficiency and reduce inter-cell interference. {\color{black}For the weighted sum rate maximization problem, its NP-hardness is proved in \cite{31,30}. An algorithmic framework combining the Lagrangian duality and dynamic programming is proposed in \cite{31} to deliver near-optimal solutions. The original problem is decomposed into two subproblems, i.e., SA and PA in \cite{30}. SA is solved using a matching algorithm, while PA is solved via geometric programming. For the energy-efficient problem, \cite{32} adopts a similar approach as \cite{30}.} 

Note that perfect CSI is assumed in the above schemes, which might be impractical for MC-NOMA systems overloaded with exceedingly number of users. Consequently, the RA under statistical CSI should be investigated. Without perfect CSI, the BS cannot decide the SIC decoding order directly, and thus, an explicit SIC decoding order should be derived first. Following this, PA and SA can be performed as for the case of perfect CSI. To further enhance the spectral efficiency of the MC-NOMA systems, full duplex (FD) BS can be introduced, achieving a substantial throughput improvement compared with FD MC-OMA and half duplex (HD) MC-NOMA systems.


The study of applying MIMO technologies to NOMA is of significance, since MIMO provides
additional degrees of freedom for further performance enhancement. {\color{black}However, the introduction of MIMO brings two major challenges: 1) it is still unclear whether MIMO-NOMA can obtain the system capacity. \cite{24} verifies that MIMO-NOMA achieves that when the users' channels are quasi-degraded. Nonetheless, the extension from quasi-degraded channels to general ones remains an open issue; 2) there exists no natural order for the users' channels in MIMO-NOMA, as they are in form of matrices or vectors. To address the issue of user ordering, an effective way is to pair users into clusters, and assign the same beamforming vectors to users in the same cluster. This decomposes the MIMO-NOMA channel into multiple separate SISO-NOMA subchannels \cite{8, capacity_zeng, 33}. A general MIMO-NOMA framework is proposed in \cite{8}, in which the inter-cluster interference is eliminated due to signal alignment-based beamforming. This further simplifies PA in MIMO-NOMA since now PA in each cluster is independent and can be treated same as in SISO. Therefore, most PA strategies for SISO, e.g., F-PA and CR-inspired PA can be directly applied. However, the above inter-cluster interference-free MIMO-NOMA framework can only be used for the case of two users per cluster. For the more general case of multiple users per cluster, inter-cluster interference generally cannot be completely removed. In this case, although the problem of channel ordering is solved by cluster-based beamforming, PA across clusters is inter-dependent, which makes the problem still non-trivial. {\color{black}A new millimeter wave transmission scheme that integrates NOMA with beamspace MIMO is proposed in \cite{33}, which shows that MIMO-NOMA can achieve higher spectrum and energy efficiency compared with existing beamspace MIMO even when there exists inter-cluster interference.}

}



{\color{black}
\section{Performance Comparison between MIMO-NOMA and MIMO-OMA}
In this section, based on the system model in \cite{8}, we conduct a performance comparison between MIMO-NOMA and MIMO-OMA when each user has a minimum rate requirement (QoS). Due to the QoS constraint, it is possible that not all users can be admitted even if the total transmit power is used. In this case, user admission is conducted to accommodate as many users as possible. On the other hand, if all users can be admitted, the objective is maximizing the sum rate of the system. For both NOMA and OMA, we consider two scenarios: 1) equal power for each cluster; and 2) cross-cluster PA. 

When equal power is allocated to each cluster, the PA for NOMA is quite simple. For each cluster, when the QoS for both users can be satisfied, PA is allocated such that the weak user satisfies its QoS, while the rest goes to the strong user to maximize the sum rate \cite{31}. Otherwise, the strong user gets all the power.
For the cross-cluster PA, first it is determined if all users can be admitted, by comparing the total power constraint with the total power required to satisfy the QoS of all users. If the required power exceeds the power constraint, the users are admitted one by one following the ascending order of required power for satisfying their QoS until all the power is consumed. Otherwise, first we assign the power to each user to ensure that its QoS is satisfied. Following this, the remaining power can be used to further increase the sum rate of the system. Since the power within each cluster should be allocated such that the weak user's QoS is satisfied, while the rest goes to the strong user, the relation between the rate increment and the extra power required only depends on the channel gain of the strongest user. Hence, the remaining power can be allocated across clusters optimally by adopting the water-filling technique only considering the channel gain of the strongest user in each cluster.  

For OMA, when equal power is allocated to each cluster, it is first determined whether both users can be admitted or not. If so, the power is allocated to the two users such that each user's QoS is satisfied, and the remaining power is then calculated. Afterwards, the water-filling technique is applied to allocate the remaining power between them. Otherwise, the strong user gets all the power. When cross-cluster PA is considered, it is first determined whether all users can be admitted or not. If so, the power is allocated to the users such that each user's QoS is satisfied, and the remaining power is calculated. Then, the water-filling technique is applied among all the users to allocate the remaining power. Otherwise, the users are arranged on descending order of their channel gains, and admitted one by one until all power is consumed.  

Simulation results for the outage probability and effective sum rate are presented in Figs. 5 and 6, respectively, in which "C-" and "E-" denote cross-cluster and equal power PA, respectively. Clearly, for both strong and weak users, C-NOMA has the lowest outage probability. Specifically, for the strong user, the outage probability for E-NOMA and E-OMA is the same, being worse than C-OMA under high transmit power. For the weak user, the outage probability for NOMA is lower than that for OMA. Moreover, cross-cluster PA has lower outage probability than equal power PA for both NOMA and OMA. For the effective sum rate, it can be seen that NOMA outperforms OMA for both the cross-cluster and equal power PA scenarios. Under low transmit power, E-NOMA achieves higher sum rate than C-NOMA due to the fact that the former allocates all power to the strong user when only the strong user can be admitted, while the latter distributes the remaining power across clusters. The same behavior can be observed for OMA. To conclude, MIMO-NOMA outperforms MIMO-OMA in terms of outage performance and effective sum rate. Moreover, optimizing the power across clusters yields significant decrease in the outage probability, as well as increase in the effective sum rate under high transmit power.  
}

\begin{figure*}
\centering
\begin{subfigure}{0.5\textwidth}
  \centering
  \includegraphics[width=1\linewidth]{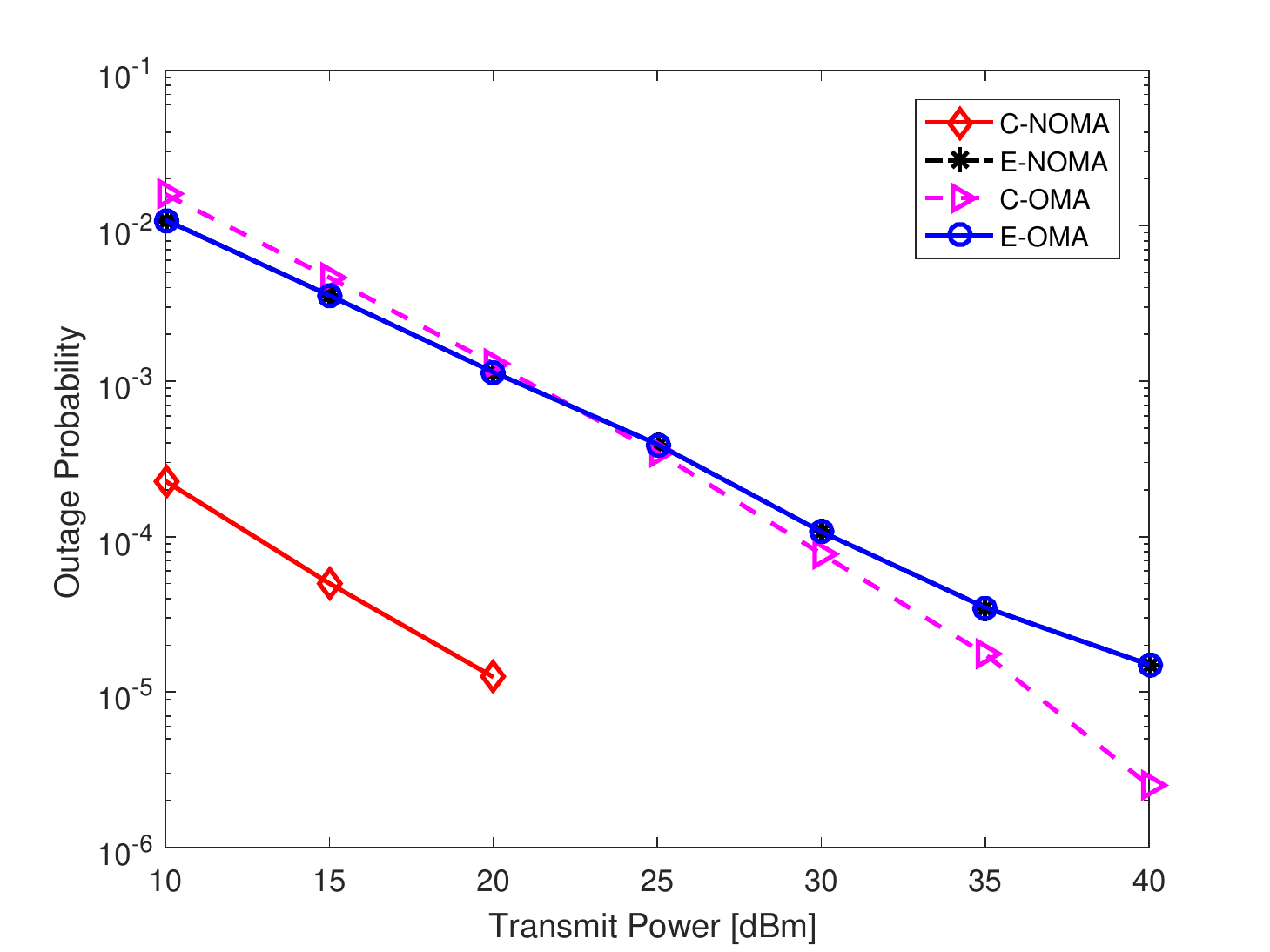}
  \caption{Strong user}
  \label{fig:sub1}
\end{subfigure}%
\begin{subfigure}{0.5\textwidth}
  \centering
  \includegraphics[width=1\linewidth]{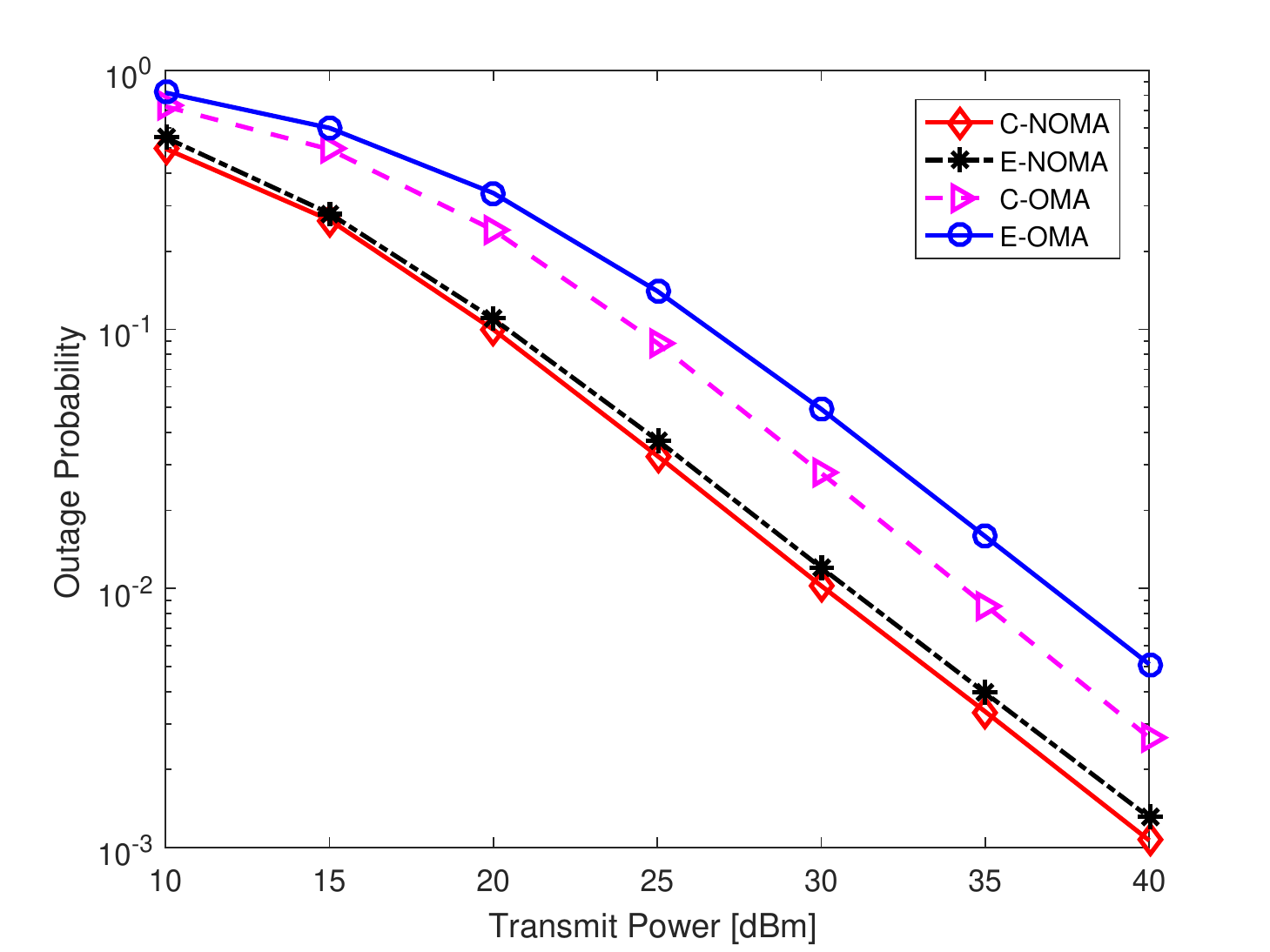}
  \caption{Weak user}
  \label{fig:sub2}
\end{subfigure}
\caption{{\color{black}Outage performance. }}
\label{fig:test}
\end{figure*}


\begin{figure}
\centering
\includegraphics[width=0.5\textwidth]{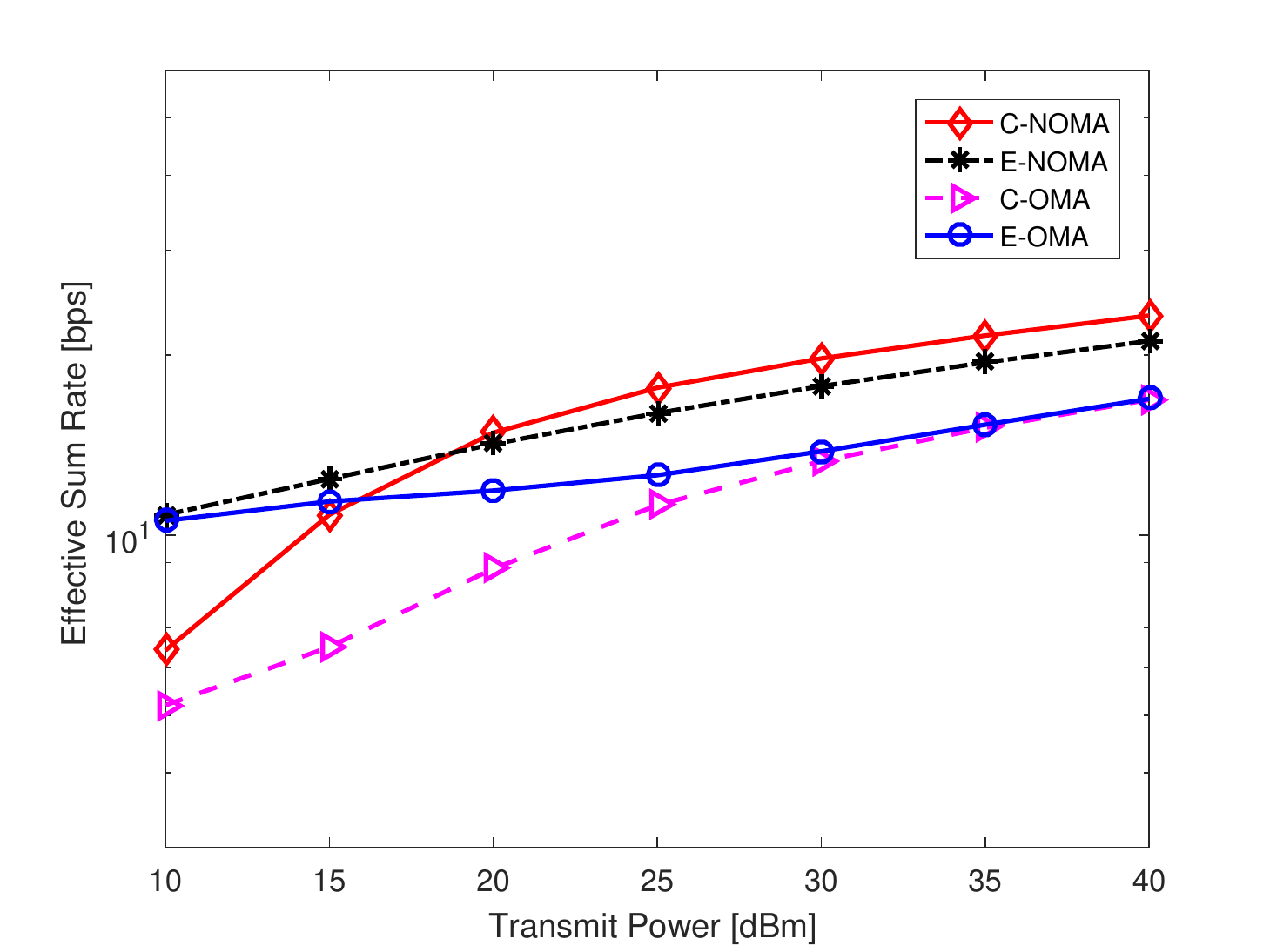}
\caption{Sum rate versus transmit power.}
\end{figure}  

\section{Practical Challenges and Research Directions}
While NOMA has attracted the attention of researchers as a potential radio access technique for 5G, the design of RA algorithms for NOMA remains still in its infancy due to various practical challenges. These include scalability, presence of inter-cell interference for multi-cell networks, integration of carrier aggregation, RA under limited channel feedback, QoS-guaranteed RA, and multi-hop communications, among others. The RA algorithms for delay-sensitive networks are inherently different from those in delay-tolerant networks as NOMA requires timely channel feedback, and further research is needed to explicitly define the RA requirements for both networks. Apart from the aforementioned challenges, one can note the following existing research gaps.

\subsection{Joint Optimization of UP and PA}
Since UP and PA are closely related to each user, a joint optimization is desirable. However, it is challenging to derive an optimal solution even under SISO, not to mention the case of MIMO. We propose D-NLUPA to ensure a fair gain among clusters for SISO. However, under the MIMO-NOMA system model in [14], D-NLUPA, NLUPA, and random paring exhibit a similar performance.  This is because under MIMO, different pairing leads to different precoding and detection matrices, which randomizes the gain of the path loss-based UP. More efforts are required to design an effective joint UP-PA strategy, especially for MIMO-NOMA.

\vspace{-0.2cm}
\subsection{RA for FD MC-MIMO-NOMA}
As research progresses, more and more advanced technologies are integrated with NOMA to
fully explore its potential, e.g., FD with MC-NOMA, MC with MIMO-NOMA and FD with
MIMO-NOMA. One can even consider these three technologies in a single framework, i.e.,
FD MC-MIMO-NOMA. This, however, yields a complex system and makes the RA problem non-trivial. Particularly, the combinatorial optimization problem of MC-NOMA is already shown
to be NP-hard. With MIMO and FD, the problem becomes much more complicated, and it is
essential to propose novel low-complexity RA schemes to ensure the superiority of NOMA over
OMA. 

\vspace{-0.2cm}
\subsection{RA for NOMA with Soft Frequency Reuse}
To solve the inter-cell interference problem, soft frequency reuse (SFR) is an indispensable element of LTE systems. Under SFR, the primary band is assigned to the cell edge users, and the secondary band is allocated to the cell center users. On the other hand, if NOMA is integrated with the SFR-based LTE systems, the signals of strong users (cell center users) and weak users (cell edge users) may be overlapped on the primary band. However, such an arrangement is unfavorable to fairness as it increases the data rate of strong users and decreases the data rate of the weak users. Additionally, the use of the primary band at the edge of the cell would generate inter-cell interference. In this regard, the design of NOMA RA for SFR-based cellular systems is an open problem.

\vspace{-0.2cm}
\subsection{Low-Complexity RA}
The existing fairness models exploit multiple parameters to adjust the fairness level, which introduces high complexity in determining the corresponding RA in NOMA. However, $\alpha$-fairness uses a single scalar, denoted by $\alpha$, to achieve different user fairness levels and well-known efficiency-fairness tradeoffs \cite{27}. The RA can be investigated for sum throughput optimization of the NOMA system with fairness constraints to obtain low-complexity algorithms.

\vspace{-0.2cm}
\subsection{Security-Aware RA}
NOMA-based communication comes with security concerns, as the user with strong channel condition needs to decode the signal of the user with weak channel condition. Thus, when the weak user becomes malicious or is under attack, the signal decoding operations of both strong and weak users are no longer reliable. For example, the attack might be in the form of an alteration of the channel quality indicator during feedback. Therefore, the design of the RA schemes becomes more critical in the presence of security concerns. To overcome this hurdle, introducing appropriate physical layer security measures is necessary, which is an interesting open problem for the research community.

\section{Concluding Remarks}
This article provides an overview of the RA algorithms for downlink NOMA in a categorized fashion. It is clear that the RA algorithms play a pivotal role in achieving the maximum benefits of NOMA. Additionally, it is shown that the fairness among the NOMA clusters can be attained by controlling the sum rate gain through the D-NLUPA algorithm. {\color{black}Moreover, for a general MIMO-NOMA system with pre-defined QoS requirement for each user, simulations are conducted, which show that MIMO-NOMA outperforms MIMO-OMA in terms of outage probability and effective sum rate when optimal PA is applied for both.} Finally, the article concludes with a discussion on open issues, including joint optimization of UP and PA, RA for FD MC MIMO-NOMA, low-complexity RA, and security-aware RA, which provides the basis for further research directions on the RA for NOMA systems.

\bibliographystyle{IEEEtran}
\bibliography{IEEEabrv,conf_short,jour_short,mybibfile}

\end{document}